\documentclass[12pt]{article}
\usepackage{amsfonts}
\usepackage{amsmath}
\usepackage{amssymb}
\usepackage{geometry}
\usepackage{apacite}
\usepackage{apalike}
\usepackage{setspace}
\usepackage{float}
\usepackage{portland}
\usepackage{scalefnt}
\usepackage{color}

\input{tcilatex}

\begin{document}

\begin{center}
\begingroup%
%TCIMACRO{\TeXButton{TeX field}{\scalefont{1.7}}}%
%BeginExpansion
\scalefont{1.7}%
%EndExpansion
\textbf{A\ Latent-Variable Bayesian Nonparametric Regression Model}\endgroup%
\medskip

\bigskip \bigskip

\textbf{George Karabatsos\medskip \footnote{%
\noindent Corresponding author. Professor, University of Illinois-Chicago,
U.S.A., Program in Measurement and Statistics. 1040 W. Harrison St. (MC\
147), Chicago, IL 60607. E-mail: gkarabatsos1@gmail.com. Phone:\
312-413-1816.}}\\[0pt]
\textit{University of Illinois-Chicago, U.S.A.}\\[0pt]
\textbf{and}\\[0pt]

\textbf{Stephen G. Walker\medskip \footnote{%
\noindent Professor, University of Kent, United Kingdom, School of
Mathematics, Statistics \& Actuarial Science. Currently, Visiting Professor
at The University of Texas at Austin, Division of Statistics and Scientific
Computation.}}\\[0pt]
\textit{University of Kent, United Kingdom}

\noindent

January 2, 2012
\end{center}

\bigskip

\noindent \textbf{Abstract:}\ We introduce a random partition model for
Bayesian nonparametric regression. The model is based on infinitely-many
disjoint regions of the range of a latent covariate-dependent Gaussian
process. Given a realization of the process, the cluster of dependent
variable responses that share a common region are assumed to arise from the
same distribution. Also, the latent Gaussian process prior allows for the
random partitions (i.e., clusters of the observations) to exhibit
dependencies among one another. The model is illustrated through the
analysis of a real data set arising from education, and through the analysis
of simulated data that were generated from complex data-generating models.%
\newline
\noindent \textbf{Keywords:} Bayesian inference; Nonparametric regression;
Gaussian process.\newpage

\section{Introduction{\textbf{\label{Introduction Section}}}}

Regression modeling is ubiquitous in many applied research areas. In
regression studies, the objective is to estimate specific distributional
aspects of a dependent variable $Y$, conditional on covariates $\mathbf{%
\mathbf{x}}=(x_{1},\ldots ,x_{p})^{\intercal }$ of interest, from a sample
data set $\mathcal{D}_{n}=\{(y_{i},\mathbf{\mathbf{x}}_{i})\}_{i=1}^{n}$,
which for notational convenience will be written as $\mathbf{X}_{n}=\mathbf{(%
\mathbf{x}}_{i}^{\intercal }\mathbf{)}_{i=1}^{n}$ and $\mathbf{y}%
=(y_{1},\ldots ,y_{n})^{\intercal }$.

Indeed, for Bayesian nonparametric regression, much research has focused on
developing random partition models (RPMs)\ that follow the general form:%
\begin{eqnarray*}
f(\mathbf{y}|\mathbf{X}_{n},\rho _{n})
&=&\dprod\limits_{d=1}^{K_{n}}\dprod\limits_{i\in S_{d}}f(y_{i}|\mathbf{x}%
_{i},\boldsymbol{\theta }_{d}) \\
\boldsymbol{\theta }_{d} &\sim &G_{0}\text{ \ \ } \\
\rho _{n} &\sim &\pi (\rho _{n}|\mathbf{X}_{n}).
\end{eqnarray*}%
In the above, $\rho _{n}=\{S_{d}\}_{d=1}^{K_{n}}$ denotes a partition of the
indices $\{1,\ldots ,n\}$ of the sample data $\{(y_{i},\mathbf{\mathbf{x}}%
_{i})\}_{i=1}^{n}$ into $K_{n}$ distinct clusters, and $\pi (\rho _{n}|%
\mathbf{X}_{n})$ denotes an RPM. These RPM's provide a very broad class of
models that encompasses product partition models (PPMs), species sampling
models (SSMs), and model-based clustering (MBC); see Quintana (2006\nocite%
{Quintana06}) for a review. A PPM\ is of the form $\pi (\rho _{n}|\mathbf{X}%
_{n})=c_{0}\tprod\nolimits_{d=1}^{K_{n}}c(S_{d}|\mathbf{X}_{n})$, with
cohesion functions $c(S_{d}|\mathbf{X}_{n})\geq 0$ (Hartigan, 1990\nocite%
{Hartigan90}; Barry \& Hartigan, 1993\nocite{BarryHartigan93}), PPMs have
been developed for Bayesian nonparametric regression (M\"{u}ller \&
Quintana, 2010\nocite{MullerQuintana10}; Park \&\ Dunson, 2010\nocite%
{ParkDunson10}; M\"{u}ller et al., 2011\nocite{MullerQuintanaRosner11}). A
SSM\ assumes the form $\pi (\rho _{n}|\mathbf{X}_{n})=\pi _{\mathbf{X}%
_{n}}(|S_{1}|,...,|S_{K_{n}}|)$ (Pitman 1996\nocite{Pitman96}; Ishwaran \&
James, 2003\nocite{IshwaranJames03}). The Dirichlet process (Ferguson, 1973%
\nocite{Ferguson73}), a popular Bayesian nonparametric model, can be
characterized either as a special PPM\ or as a special SSM. On the other
hand, with MBC, a random partition $\rho
_{n}=\{S_{d}=\{i:d_{i}=d\}\}_{d=1}^{K_{n}}$ is formed by sampling latent
indicators $d_{i}$, $(i=1,\ldots ,n)$, from weights $\omega _{j}$ of a
discrete mixture model.

In this paper, we develop and illustrate a novel Bayesian nonparametric
regression model, which may be characterized as an RPM. The model randomly
partitions the $n$ observations into distinct clusters that each share a
common region of a transformed\ covariate space, and then given the
covariate $\mathbf{\mathbf{x}}$, uses the dependent responses in the
covariate region to predict $Y$. Specifically, the novel regression model is
based on a fixed partition $(A_{j})$ of the range $%
%TCIMACRO{\U{211d} }%
%BeginExpansion
\mathbb{R}
%EndExpansion
=$ $\cup _{j=-\infty }^{\infty }(A_{j}=(j-1,j])$ and a Gaussian Process (GP) 
$z(\mathbf{x})$ which induces a random partition by $\rho _{n}=\{S_{d}=\{i:z(%
\mathbf{x}_{i})\in A_{d}\}\}_{d=1}^{K_{n}}.$

To further elaborate, consider the standard Bayesian nonparametric mixture
model, with latent variable for the component. Such a model is given by 
\begin{equation*}
f(y,d)=w_{d}\,f(y|\theta _{d}).
\end{equation*}%
The $d$ classifies which component $f(y|\theta )$ the observation $y$ comes
from and the weight $w_{d}$ is the population probability of coming from
component $d$. There has been much debate and proposals as to how covariates 
$\mathbf{x}$ enter into such a model in a meaningful way. Following the RPM
idea it makes most sense that if $\mathbf{x}$ and $\mathbf{x}^{\prime }$ are
close then observations $y$ and $y^{\prime }$ would be expected to come from
the same component. Hence, it is appropriate to make the $d$ depend on $%
\mathbf{x}$. A convenient way to achieve this is via a Gaussian process $z(%
\mathbf{x})$, such that%
\begin{equation*}
d(\mathbf{x})=j\iff z(\mathbf{x})\in A_{j}
\end{equation*}%
where $(A_{j})$ is a fixed partition of $\mathbb{R}$, i.e. $\cup _{j}A_{j}=%
\mathbb{R}$ and $A_{j}\cap A_{j^{\prime }}=\emptyset $ for $j\neq j^{\prime
} $.

The usual idea of having the weights depend on $\mathbf{x}$ in the form $%
\omega _{j}(\mathbf{x})$ and having $\omega _{j}(\mathbf{x})$ close to $%
\omega _{j}(\mathbf{x}^{\prime })$ whenever $\mathbf{x}$ is close to $%
\mathbf{x}^{\prime }$, is a rather weak condition. While in this case the
densities for $y$ and $y^{\prime }$ may be close to each other, there is no
suggestion that $y$ and $y^{\prime }$ are coming from the same component,
which is the more realistic notion. So what is needed is to have $y$ close
to $y^{\prime }$ in probability, rather than simply close in distribution.

Therefore, the proposed model is given by 
\begin{equation*}
f(y,d|z,\mathbf{x})=\mathbf{1}(z(\mathbf{x})\in A_{d})\,f(y|\theta _{d}).
\end{equation*}%
So 
\begin{equation*}
f(y|z,\mathbf{x})=\sum_{j}\mathbf{1}(z(\mathbf{x})\in A_{j})\,f(y|\theta
_{j})
\end{equation*}%
and 
\begin{equation*}
f(y|\mathbf{x})=\sum_{j}\omega _{j}(\mathbf{x})\,f(y|\theta _{j})
\end{equation*}%
where 
\begin{equation*}
\omega _{j}(\mathbf{x})=P(z(\mathbf{x})\in A_{j}).
\end{equation*}%
In Karabatsos and Walker (2012\nocite{KarabatsosWalker12c}), this model was
employed where $z(\mathbf{x})\sim \mathrm{n}(\eta (\mathbf{x}),\sigma ^{2}(%
\mathbf{x}))$. It was explained in that paper how $\sigma (\mathbf{x})$
controlled the modes of $f(y|\mathbf{x})$ and why this was an important
aspect of the model in keeping with the idea that $\mathbf{x}$ close to $%
\mathbf{x}^{\prime }$ determines $y$ and $y^{\prime }$ coming from the same
component. That is, for $\mathbf{x}$ close to $\mathbf{x}^{\prime }$, it can
be that $\omega _{j}(\mathbf{x})$ and $\omega _{j}(\mathbf{x}^{\prime })$
are both close to 1 for some $j$. Henceforth, we refer to Karabatsos and
Walker's (2012\nocite{KarabatsosWalker12c}) model as the "independence
model", because it assumes independent latent variables $z(\mathbf{x})\sim 
\mathrm{n}(\eta (\mathbf{x}),\sigma ^{2}(\mathbf{x}))$ and $z(\mathbf{x}%
^{\prime })\sim \mathrm{n}(\eta (\mathbf{x}^{\prime }),\sigma ^{2}(\mathbf{x}%
^{\prime }))$ for any two distinct covariates $\mathbf{x}$ and $\mathbf{x}%
^{\prime }$.

In the present paper we acknowledge that it would be further desirable for
the $z$ process to be constructed with dependence; i.e. 
\begin{equation*}
z(\cdot )\sim \mathrm{GP}\left[ \eta (\cdot ),\sigma (\cdot ,\cdot )\right] .
\end{equation*}%
This will reinforce the notion that it is required for $\omega _{j}(\mathbf{x%
})$ and $\omega _{j}(\mathbf{x}^{\prime })$ to both be close to 1 when $%
\mathbf{x}$ is close to $\mathbf{x}^{\prime }$. The dependent Gaussian
process facilitates this to a greater extent than under the independent
process.

In terms of a RPM, we have 
\begin{equation*}
P\left( d_{1},\ldots ,d_{n}|\mathbf{x}_{1},\ldots ,\mathbf{x}_{n}\right)
=P\left( z(\mathbf{x}_{1})\in A_{d_{1}},\ldots ,z(\mathbf{x}_{n})\in
A_{d_{n}}\right) .
\end{equation*}%
This is an appealing version of a probability for the partition as it
marginalizes from higher to lower dimensions, addressing the curse of
dimensionality. Also, it is clear that since our model allows for the GP\ to
exhibit dependencies among the latent variables $z(\mathbf{x}_{1})\in
A_{d_{1}},\ldots ,z(\mathbf{x}_{n})\in A_{d_{n}}$, it is in a sense\ more
flexible than a PPM\ because it does not force partitions under the
assumption that $\pi (\rho _{n}|\mathbf{X}_{n})$ is a product prior.\noindent

\begin{center}
--- \ Insert \noindent Figure 1 \ ---
\end{center}

Figure 1 illustrates the mixture weights $\omega _{j}(\mathbf{x})$ and the
resulting predictive densities $f(y|\mathbf{x})$ of the model, for a single
covariate $\mathbf{x}=x$ having observed values $x_{1}=1,$ $x_{2}=1.3,$ and $%
x_{2}=4$. Also, the figure assumes $\eta (x_{1})=-.30$, $\eta (x_{2})=.21$, $%
\eta (x_{3})=4.8$, and the squared-exponential covariance function $\sigma
(x,x^{\prime })=\sigma _{C}^{2}\exp (-.5||x-x^{\prime }||^{2})$, and
presents the weights and the densities for small $\sigma _{C}^{2}=.01$ and
for large $\sigma _{C}^{2}=10$. Throughout, $||\cdot ||$ denotes the
Euclidean norm. As shown, when either $\sigma _{C}^{2}$ is small or large,
the mixture weights $\omega _{j}(x)$ and the resulting predictive densities $%
f(y|x)$ are similar when $x$ and $x^{\prime }$ are close. The weights and
densities become more dissimilar as the distance between $x$ and $x^{\prime }
$ increases. Also, the parameter $\sigma _{C}^{2}$ controls the number of
modes in $f(y|\mathbf{x})$. At one extreme, as $\sigma _{C}^{2}$ decreases, $%
f(y|\mathbf{x})$ becomes more unimodal. As $\sigma _{C}^{2}$ increases, $f(y|%
\mathbf{x})$ becomes more multimodal.

We now describe the layout of the rest of the paper. In Section \ref{The
Regression Model section} we fully present our regression model. In Section %
\ref{Illustrations Section}, we illustrate our model through the analysis of
real and simulated data sets. In so doing, we compare the predictive
performance of our new model, against the previous version of our regression
model which assumes independent latent variables $z(\mathbf{x}_{1}),\ldots
,z(\mathbf{x}_{n}),$ and against another regression model that is known to
provide good predictive performance. Section \ref{Conclusion Section}
concludes with a discussion.

\section{The Regression Model{\textbf{\label{The Regression Model section}}}}

For a sample set of data $\mathcal{D}_{n}=\{(\mathbf{x}_{i},y_{i})%
\}_{i=1}^{n}$, our Bayesian nonparametric regression model has parameters $%
\boldsymbol{\zeta }=((\boldsymbol{\theta }_{j})_{j\in \mathbb{Z}},%
\boldsymbol{\beta },\sigma _{\mathcal{C}}^{2},\boldsymbol{\phi })$, along
with latent indicator parameters $\mathbf{d}=(d_{1},\ldots
,d_{n})^{\intercal }$. The model is defined by: 
\begin{subequations}
\label{NewModel}
\begin{eqnarray}
f(\mathbf{y},\mathbf{d}|\mathbf{X}_{n},z,\boldsymbol{\zeta })
&=&\dprod\limits_{i=1}^{n}f(y_{i}|\boldsymbol{\theta }_{d_{i}})\mathbf{1}(z(%
\mathbf{x}_{i})\in A_{d_{i}}),\text{ }i=1,\ldots ,n,  \label{OrigLike} \\
\pi (\boldsymbol{\theta }) &=&\dprod\limits_{j=-\infty }^{\infty }\pi _{j}(%
\boldsymbol{\theta }_{j}), \\
(z(\mathbf{x}_{1}),\ldots ,z(\mathbf{x}_{n})) &\sim &\text{\textrm{n}}_{n}(%
\mathbf{X}_{1n}\boldsymbol{\beta },\sigma _{\mathcal{C}}^{2}(\mathcal{C}_{%
\boldsymbol{\phi }}(\mathbf{x}_{i},\mathbf{x}_{l}))_{n\times n}), \\
\boldsymbol{\beta },\sigma _{\mathcal{C}}^{2},\boldsymbol{\phi } &\sim &%
\mathrm{n}_{p+1}(\boldsymbol{\beta }|\mathbf{m}_{\boldsymbol{\beta }},\sigma
_{\mathcal{C}}^{2}\mathbf{\Sigma }_{\boldsymbol{\beta }})\mathrm{ga}(\sigma
_{\mathcal{C}}^{-2}|a_{\mathcal{C}},b_{\mathcal{C}})\pi (\boldsymbol{\phi })
\end{eqnarray}%
where $\mathbf{X}_{1n}=(1,\mathbf{x}_{i}^{\intercal })_{n\times (p+1)}$,
while $\mathrm{n}_{_{K}}(\boldsymbol{\cdot }|\boldsymbol{\cdot },\boldsymbol{%
\cdot })$ and $\mathrm{ga}(\boldsymbol{\cdot }|\boldsymbol{\cdot },%
\boldsymbol{\cdot })$ are respectively the probability density functions of
the $K$-variate normal and gamma distributions (shape and rate
parameterized). As shown, the model is based on a GP, with mean function $%
\mathbf{X}_{1n}\boldsymbol{\beta }$ and covariance function matrix $\sigma _{%
\mathcal{C}}^{2}(\mathcal{C}_{\boldsymbol{\phi }}(\mathbf{x}_{i},\mathbf{x}%
_{l}))_{n\times n}$, where $\boldsymbol{\beta }$ $=(\beta _{0},\beta
_{1},\ldots ,\beta _{p})^{\intercal }$, and where $\mathcal{C}_{\boldsymbol{%
\phi }}(\mathbf{\cdot },\mathbf{\cdot })$ is a correlation function that
depends on the parameter $\boldsymbol{\phi }$.

A standard choice of kernel densities is provided by univariate normal
densities $f(\cdot |\boldsymbol{\theta }_{j})=$ \textrm{n}$(\cdot |\mu
_{j},\sigma _{j}^{2})$ ($j=0,\pm 1,\pm 2,\ldots $), which may be assigned
conjugate prior density: 
\end{subequations}
\begin{equation*}
\pi (\boldsymbol{\theta })=\pi (\boldsymbol{\mu },\boldsymbol{\sigma }%
^{2})=\dprod\limits_{j=-\infty }^{\infty }\mathrm{n}(\mu _{j}|\mu _{\mu
j},\sigma _{\mu j}^{2})\mathrm{ga}(\sigma _{j}^{-2}|\alpha _{\sigma j},\beta
_{\sigma j}).
\end{equation*}%
For the covariance function $\sigma _{\mathcal{C}}^{2}\mathcal{C}_{%
\boldsymbol{\phi }}(\cdot ,\cdot )$, possible choices of the correlation
function include the powered-exponential family $\mathcal{C}_{\boldsymbol{%
\phi }}(\mathbf{x},\mathbf{x}^{\prime })=\exp (-\phi _{1}||\mathbf{x}-%
\mathbf{x}^{\prime }||^{\phi _{2}})$ (for $\phi _{1}>0$; $0<\phi _{2}\leq 2$%
), the Cauchy family, the Mat\'{e}rn family, as well as families of
correlation functions that are either non-stationary or non-isotropic (e.g.,
Rasmussen \&\ Williams, 2006\nocite{RasmussenWilliams06}).

Given data $\mathcal{D}_{n}$ likelihood $\tprod\nolimits_{i=1}^{n}f(y_{i}|%
\mathbf{x}_{i};\boldsymbol{\zeta })$, with $f(y_{i}|\mathbf{x}_{i};%
\boldsymbol{\zeta })=\sum_{j}f(y|\theta _{j})\omega _{j}(\mathbf{x;}%
\boldsymbol{\beta },\sigma _{\mathcal{C}}^{2},\boldsymbol{\phi })$ (see
Section 1), and a proper prior density $\pi (\boldsymbol{\zeta })$ defined
over $\Omega _{\boldsymbol{\zeta }}=\{\boldsymbol{\zeta }\}$, the posterior
density of $\boldsymbol{\zeta }$ is proper and given by:%
\begin{equation*}
\pi (\boldsymbol{\zeta }|\mathcal{D}_{n})\propto
\dprod\limits_{i=1}^{n}f(y_{i}|\mathbf{x}_{i};\boldsymbol{\zeta })\pi (%
\boldsymbol{\zeta })
\end{equation*}%
up to a proportionality constant.\ Then the posterior predictive density of $%
Y$\ is defined by:%
\begin{equation*}
f_{n}(y|\mathbf{x})=\int f(y|\mathbf{x};\boldsymbol{\zeta })\pi (\boldsymbol{%
\zeta }|\mathcal{D}_{n})\mathrm{d}\boldsymbol{\zeta },
\end{equation*}%
with this density corresponding to posterior predictive mean and variance%
\begin{equation*}
\mathrm{E}_{n}(Y_{i}|\mathbf{x}_{i})=\tint yf_{n}(y|\mathbf{x}_{i})\mathrm{d}%
y;\text{ \ \ }\mathrm{Var}_{n}(Y_{i}|\mathbf{x}_{i})=\tint \{y-\mathrm{E}%
(Y_{i}|\mathbf{x}_{i})\}^{2}f_{n}(y|\mathbf{x}_{i})\mathrm{d}y.
\end{equation*}%
In the present paper, in applications of our regression model, our emphasis
is in prediction rather than inference of the model parameters $\boldsymbol{%
\zeta }$. Hence, we focus statistical inferences on the posterior predictive
density $f_{n}(y|\mathbf{x})$, and functionals of interest.

The posterior densities $\pi (\boldsymbol{\zeta }|\mathcal{D}_{n})$ and $%
f_{n}(y|\mathbf{x})$\ can be estimated by using standard Gibbs MCMC\
sampling methods for infinite-dimensional models, which make use of
strategic latent variables (Kalli, Griffin, \&\ Walker, 2010\nocite%
{KalliGriffinWalker10}). The Appendix provides more details. Also the
Appendix describes how the model and corresponding MCMC methods can be
easily extended to handle the analysis of censored observations, discrete
dependent variables, and the analysis of spatial or spatio-temporal data via
an appropriate modification of the GP\ covariance function.

\subsection{Model Assessment and Comparison Methods}

After $M$\ regression models are fit to a data set $\mathcal{D}_{n}$, the
predictive performance of each model $m\in \{1,\ldots ,M\}$ can be assessed
by the mean-square predictive-error criterion 
\begin{equation}
D(m)=\sum_{i=1}^{n}\{y_{i}-\text{\textrm{E}}_{n}(Y_{i}|\mathbf{x}%
_{i},m)\}^{2}+\sum_{i=1}^{n}\mathrm{Var}_{n}(Y_{i}|\mathbf{x}_{i},m)
\label{Pred Criterion}
\end{equation}%
(Gelfand \& Ghosh, 1998\nocite{GelfandGhosh98}). The criterion is often used
in the practice in the assessment and comparison of Bayesian models (e.g.,
Gelfand \& Banerjee, 2010\nocite{GelfandBanerjee10}). The first term of (\ref%
{Pred Criterion}) measures data goodness-of-fit, and the second term is a
penalty that is large for models which either over-fit or under-fit the
data. The criterion (\ref{Pred Criterion}) can be re-written as%
\begin{equation*}
D(m)=\sum_{i=1}^{n}\dint (y-y_{i})^{2}f_{n}(y|\mathbf{x}_{i},m)\mathrm{d}%
y=\sum_{i=1}^{n}\text{\textrm{E}}_{n}[(y_{i}-y)^{2}|\mathbf{x}_{i},m].
\end{equation*}%
So the estimate of $D(m)$ is obtained by generating posterior predictive
samples $y_{i}^{\mathrm{pred}(s)}|\mathbf{x}_{i}$ ($i=1,\ldots ,n$) at each
iteration $s=1,\ldots ,S\ $of the MCMC\ chain, and then taking%
\begin{equation*}
\widehat{D}(m)=\dfrac{1}{S}\sum_{s=1}^{S}\sum_{i=1}^{n}\left( y_{i}-y_{i}^{%
\mathrm{pred}(s)}\right) ^{2}=\sum_{i=1}^{n}\widehat{D}_{i}(m),
\end{equation*}%
where the individual quantities $\widehat{D}_{i}(m)$ can be used to provide
a more detailed assessment about a model's predictive performance. The
Appendix provides some more details about the MCMC\ methods for estimating $%
D(m)$.

\section{Illustrations\label{Illustrations Section}}

\subsection{\textit{Math Teaching Data}}

Here we illustrate the proposed model of equation (\ref{NewModel}), through
the analysis of data that were collected to study a new undergraduate
teacher education curriculum, instituted in 2009 by four Chicago-area
universities. The study aimed to evaluate the impact of the new curriculum
on the ability to teach math among $n=89$ of its second-year students.
Impact is measured by a dependent variable called "change" (mean=$.80$; s.d.=%
$.6$), which is the change in math teaching ability score of the student,
from before (pre-test) and after (post-test) completing a course in math
teaching. Also, there are three covariates. The first covariate is lmt140,
where lmt140=1 if the course is learning of math teaching (lmt)\ level 140,
and lmt140=0 if the course is lmt141 (mean=$.73$; s.d.=$.6$). The second
covariate is uic, which is a 0-1 indicator of whether the student is from
the University of Illinois-Chicago, versus one of the other three
universities (mean=$.60$; s.d.=$.5$). The third covariate is pretest score
(mean=$-.83$; s.d.=$.8$). Each of the three covariates were z-standardized
to have mean zero and variance 1, prior to data analysis.

For the regression model presented in equation (\ref{NewModel}), we assumed
a squared-exponential covariance function for the GP, given by $\sigma _{%
\mathcal{C}}^{2}\mathcal{C}_{\boldsymbol{\phi }}(\mathbf{x},\mathbf{x}%
^{\prime })=\sigma _{\mathcal{C}}^{2}\exp (-.5||\mathbf{x}-\mathbf{x}%
^{\prime }||^{2})$. Also for this model we assigned mostly high-variance
priors $\mu _{j}\sim _{iid}\mathrm{n}(\mu _{\mu }=0,\sigma _{\mu }^{2}=10)$, 
$\sigma _{j}^{-2}\sim _{iid}\mathrm{ga}(1,10^{-3})$, $\boldsymbol{\beta }%
|\sigma _{\mathcal{C}}^{2}\sim \mathrm{n}(\mathbf{0},\sigma _{\mathcal{C}%
}^{2}10^{5}\mathbf{I}_{p+1})$, and $\sigma _{\mathcal{C}}^{-2}\sim \mathrm{ga%
}(1,10^{4})$, to reflect the relative lack of prior information about these
parameters. The gamma prior for $\sigma _{\mathcal{C}}^{2}$ reflects our
prior belief that the conditional density of the change score, $f(y|\mathbf{x%
})$, tends to be unimodal. This implies the belief that the covariates $%
\mathbf{x}$ tend to be informative about the dependent variable. To estimate
the model, a total of 150,000 sampling iterations of the MCMC\ algorithm
were run (see Appendix), and the last 75,000 samples were used to estimate
the model's posterior distribution (after burn-in). The posterior predictive
samples and the $D(m)$ criterion stabilized over the MCMC iterations, and
the resulting posterior predictive and $\widehat{D}(m)$ estimates had
near-zero 95\%\ Monte Carlo confidence intervals (MCCI) according to a
consistent batch means estimator (Jones et al., 2006\nocite{Jones_etal06}).

\begin{center}
--- \ Insert Figures 2 and 3\ ---
\end{center}

Figure 2 presents the posterior predictive mean and variance estimates of
the change dependent variable, conditional on the three covariates (lmt140,
uic, pretest). The relation between change score and pretest score is quite
nonlinear. Figure 3 presents the posterior predictive density estimates of
the change score, for a range of values of the pretest scores, while
conditioning on uic=1 and lmt140=1. These densities are shown to be skewed
and unimodal. In summary, the result show that the new teacher education
curriculum tended to have a positive effect on mathematics teaching ability,
over time.

We also analyzed the data using a simpler version of our regression model (%
\ref{NewModel}), namely the "independence model" (see Section \ref%
{Introduction Section}), for which we specified $z(\mathbf{x}_{i})\sim
_{ind} $ \textrm{n}$((1,\mathbf{x}_{i}^{\mathbf{\intercal }})\boldsymbol{%
\beta },\sigma _{\mathcal{C}}^{2}(\mathbf{x}_{i}))$, with $\sigma _{\mathcal{%
C}}^{2}(\mathbf{x})=\exp ((1,\mathbf{x}_{i}^{\mathbf{\intercal }})%
\boldsymbol{\lambda })$, along with priors $\mu _{j}\sim _{iid}\mathrm{n}%
(0,10)$, $\sigma _{j}^{-2}\sim _{iid}\mathrm{ga}(1,10^{-3})$, and $(%
\boldsymbol{\beta },\boldsymbol{\lambda })\sim \mathrm{n}(\mathbf{0},10^{5}%
\mathbf{I}_{2(p+1)})$. Thus, this independence model assumed the same priors
for $(\boldsymbol{\mu },\boldsymbol{\sigma }^{2})$, as in the GP-based model
described earlier. A previous study (Karabatsos \& Walker, 2012\nocite%
{KarabatsosWalker12c}) showed that the independence model tended to have
better predictive performance than 26 other regression models (according to
the $D(m)$ criterion), over many real and simulated data sets, with the
BART\ model (Bayesian Additive Regression Trees model; Chipman, et al. 
\nocite{ChipmanGeorgeMcCulloch10}2010) being among the more competitive
models. For the data set under current consideration, the independence model
was estimated by 150,000 MCMC\ sampling iterations, after discarding the
first 75,000 samples (burn-in) (see Appendix A of Karabatsos \& Walker, 2012%
\nocite{KarabatsosWalker12c}). Also, the BART\ model was fit to the data
set, via the generation of 42,000 posterior samples via a Bayesian
back-fitting algorithm. For each of these two models, the samples of the $%
D(m)$ criterion stabilized over sampling iterations, and the resulting $%
\widehat{D}(m)$ estimate had a small 95\% MCCI.

For the current data set of the $n=89$ students under the new teacher
education curriculum, our GP-based regression model (\ref{NewModel}) had a
better predictive performance ($D(m)=1.3$; \ MCCI=$\pm .1$)) than the
independence model ($D(m)=5.1$; \ MCCI=$\pm .8$)), and the BART\ model ($%
D(m)=52.5$; \ MCCI=$\pm .04$)) which was fit by implementing the BayesTree
package of the R statistical software (Chipman \&\ McCulloch, 2010\nocite%
{ChipmanMcCulloch10}). Also, for the GP-based model, had no outliers, as the 
$\widehat{D}_{i}(m)$ estimates over the $n=89$ observations had a 5-number
summary (i.e., min, 25\%ile, 50\%ile, 75\%ile, and max) of $%
\{.00,.01,.01,.02,.11\}$. For the independence model, it was $%
\{.01,.02,.03,.06,.70\}$. In summary, it seems that the predictive accuracy
of the regression model can be substantially improved by accounting for
dependence among the latent variables $(z(\mathbf{x}_{1}),\ldots ,z(\mathbf{x%
}_{n}))$. In the next subsection, we use a simulation study to further
investigate this issue.

\subsection{{{\textit{Complex Regression Functions}}}}

Here, using a range of complex data-generating models, we conduct a
simulation study to compare the predictive performance between the GP-based
regression model, the independence model, and the BART model. They include
data-generating models where $f(y|\mathbf{x})$ is a unimodal sampling
density, with mean depending on complex functions of $\mathbf{x}$. They also
include data-generating models where $f(y|\mathbf{x})$ is a multimodal
sampling density, having mean and number of modes that also depend on
complex functions of $\mathbf{x}$.

For the unimodal $f(y|\mathbf{x})$\ setting, we consider two data-generating
models which respectively assumed the following mean functions for the
dependent variable:%
\begin{eqnarray}
\mathrm{E}_{1}(Y|\mathbf{x}) &=&1.9[1.35+\exp (x_{1})\sin
(13(x_{1}-.6)^{2})\exp (-x_{2})\sin (7x_{2})],  \label{Hwang} \\
\text{\textrm{E}}_{4}(Y|\mathbf{x}) &=&10\sin (\pi
x_{1}x_{2})+20(x_{3}-.5)^{2}+10x_{4}+5x_{5}+\tsum\nolimits_{k=6}^{10}0x_{k}.
\label{Friedman}
\end{eqnarray}%
Equation (\ref{Hwang}) is a complex 2-dimensional covariate interaction
(Hwang, et al., 1994\nocite{HwangLayetal94}). Equation (\ref{Friedman}) is a
complex function of ten covariates, with 5 covariates irrelevant (Friedman,
1991\nocite{Friedman91}). With respect to these two functions, we generated
a data set of $n=225$ observations from $\mathrm{n}(y_{i}|\mathrm{E}_{1}(Y|%
\mathbf{x}_{i}),.0625)\mathrm{u}_{2}(\mathbf{x}_{i}|0,1)$, and we generated
another data set of $n=100$ observations from $\mathrm{n}(y_{i}|\mathrm{E}%
_{4}(Y|\mathbf{x}_{i}),\sigma _{i}^{2})\mathrm{u}_{10}(\mathbf{x}_{i}|0,1),$
for $i=1,\ldots ,n$, where $\mathrm{u}_{K}(\mathbf{x}_{i}|0,1)$ denotes the
density function of a $K$-variate uniform distribution.

We simulated two additional data sets under settings where $f(y|\mathbf{x})$
is multimodal and based on mixtures of normal densities, 10 covariates $%
\mathbf{x}$, and with the number of modes depending on $\mathbf{x}$.
Specifically, the number of modes $N_{\func{mod}}(\mathbf{x})$ in the
density $f(y|\mathbf{x})$ ranged from 1 to 4, via the function $N_{\func{mod}%
}(\mathbf{x})=\min (\max (\mathrm{floor}(\mathrm{E}_{5}(Y|\mathbf{x})),1),4)$%
, with \textrm{E}$_{5}(Y|\mathbf{x})=(3,1.5,0,0,2,0,0,0)(x_{1},\ldots
,x_{8})^{\mathbf{\intercal }}$. The four modes are respectively defined by $%
\mathrm{E}_{1}(Y|\mathbf{x}),$ \textrm{E}$_{2}(Y|\mathbf{x})$, \textrm{E}$%
_{3}(Y|\mathbf{x})$, and \textrm{E}$_{4}(Y|\mathbf{x})$, with $\mathrm{E}%
_{1}(Y|\mathbf{x})$ and \textrm{E}$_{4}(Y|\mathbf{x})$ given by (\ref{Hwang}%
) and (\ref{Friedman}), along with\noindent 
\begin{eqnarray*}
\text{\textrm{E}}_{2}(Y|\mathbf{x}) &=&(-2x_{1})^{\mathbb{I}%
(x_{1}<.6)}(-1.2x_{1})^{\mathbb{I}(x_{1}\geq .6)}+\cos (5\pi
x_{2})/(1+3x_{2}^{2}), \\
\text{\textrm{E}}_{3}(Y|\mathbf{x}) &=&((x_{1},x_{2},x_{3},x_{4})\boldsymbol{%
\beta })^{2}\exp ((x_{1},x_{2},x_{3},x_{4})\boldsymbol{\beta });\boldsymbol{%
\beta }=(2,1,1,1)^{\top }/7^{1/2}.
\end{eqnarray*}%
We simulated a data set of $n=100$ observations from a sampling density $%
\mathrm{n}(y_{i}|\mathrm{E}_{d_{i}}(Y|\mathbf{x}_{i}),\sigma _{i}^{2})$ $%
\times \mathrm{u}_{10}(\mathbf{x}_{i}|0,1)$, for $i=1,\ldots ,n$, with $%
d_{i}\sim \mathrm{u}\{1\}$ when $N_{\func{mod}}(\mathbf{x}_{i})=1$, $%
d_{i}\sim \mathrm{u}\{1,2\}$ when $N_{\func{mod}}(\mathbf{x}_{i})=2$, $%
d_{i}\sim \mathrm{u}\{1,2,3\}$ when $N_{\func{mod}}(\mathbf{x}_{i})=3$, and $%
d_{i}\sim \mathrm{u}\{1,2,3,4\}$ when $N_{\func{mod}}(\mathbf{x}_{i})=4$.
Also, we simulated another data set, of $n=225$ observations, from the same
density.\noindent \noindent

\begin{center}
--- \ Insert Table \ref{Sim Table} \ ---
\end{center}

To analyze each of the four simulated data sets described in this
subsection, our GP-based regression model, and our independence model, each
assumed priors $\mu _{j}\sim _{i.i.d.}$\textrm{n}$(\widehat{\mu },100)$,
with $\widehat{\mu }$ the empirical mean of the simulated $Y$. Otherwise,
each of these models assumed the same priors for their other parameters, and
the GP-based model assumed the same squared-exponential covariance function
for z-standardized covariates, as in the previous subsection. Moreover, each
of these two models were estimated according to 150,000 MCMC\ sampling
iterations, after discarding the first 75,000 samples (burn-in). Also, the
BART\ model was fit to each of the four data sets, via the generation of
300,000 posterior samples. For each of the three models, the $D(m)$
criterion stabilized over MCMC iterations, and the resulting $\widehat{D}(m)$
estimate yielded a small 95\%\ MCCI.

Table \ref{Sim Table} presents the results of the simulation study, in the
comparison of the three models, in terms of the mean-squared predictive
error $D(m)$. The GP-based model obtained the best $D(m)$ predictive
performance for all the simulated data sets. Also, for the GP-based model,
the individual $\widehat{D}_{i}(m)$ predictive errors tended to be quite
small, even though the true data-generating models were quite complex.
Specifically, for the 2-dimensional unimodal, 10-dimensional unimodal, the
multimodal ($n=100$), and the multimodal ($n=225$) simulated data sets, the
model obtained $\widehat{D}_{i}(m)$ 5-number summaries (i.e., min, 25\%ile,
50\%ile, 75\%ile, and max) of $\{.01,.01,.02,.02,.15\}$, $%
\{.01,.02,.03,.04,.10\}$, $\{.00,.02,.02,.03,.07\}$, and $%
\{.01,.02,.02,.04,4.8\}$, respectively.

\section{Conclusions\label{Conclusion Section}}

We have described a Bayesian nonparametric regression model, and
demonstrated the suitability of the model through the analysis of both real
and simulated data sets. The key idea of the paper is that close covariates $%
x$ and $x^{\prime }$ should result in $y$ and $y^{\prime }$ being close in
probability, rather than in distribution, which has led to the current
prevailing model constructions. Close in probability suggest outcomes from
close covariates share a common component distribution which is, in our
case, modeled as a normal distribution. For this to happen the weights at a
particular component value for these similar covariates should both be close
to 1, and to facilitate this a dependent Gaussian process is the most
suitable model. Hence, all the aspects of the model play a clear discernible
role.

\bigskip \bigskip

\noindent \begingroup%
%TCIMACRO{\TeXButton{TeX field}{\bf\Large}}%
%BeginExpansion
\bf\Large%
%EndExpansion
\noindent {\Large \noindent }Appendix: MCMC Algorithm\endgroup

Our infinite-dimensional regression model can be estimated via the
implementation of the MCMC sampling methods of Kalli et al. (2010\nocite%
{KalliGriffinWalker10}). This method involves introducing strategic latent
variables, to implement exact MCMC algorithms for the estimation of the
model's posterior distribution. Specifically, for our regression model
(Section \ref{The Regression Model section}), we introduce new latent
variables $(u_{i})_{i=1}^{n}$, and a decreasing function $\xi _{d}=\exp
(-|d|)$, such that the model's data likelihood can be rewritten as the joint
distribution: 
\begin{equation}
\dprod\limits_{i=1}^{n}f(y_{i},d_{i},u_{i}|\mathbf{x}_{i},z)=\dprod%
\limits_{i=1}^{n}\{\mathbf{1}(0<u_{i}<\xi _{d_{i}})\xi _{d_{i}}^{-1}f(y_{i}|%
\boldsymbol{\theta }_{d_{i}})\mathbf{1}(z(\mathbf{x}_{i})\in A_{d_{i}})\}.
\label{Latlike norm}
\end{equation}%
Marginalizing over the latent variables $u_{i}$ in (\ref{Latlike norm}), for
each $i=1,\ldots ,n$, returns the original model (eq. \ref{OrigLike}). Thus,
given the new latent variables, the infinite-dimensional model can be
treated as a finite-dimensional model. This, in turn, permits the use of
standard MCMC methods to sample the model's full joint posterior
distribution. Given all variables, save the $(d_{i})_{i=1}^{n}$, the choice
of each $d_{i}$\ have minimum $-N_{\max }$ and maximum $N_{\max }$, where $%
N_{\max }=\max_{i}[\max_{j}\mathbf{1}(u_{i}<\xi _{j})|j|]$.

Then for our regression model, assuming the normal kernel densities $f(y_{i}|%
\boldsymbol{\theta }_{j})=\mathrm{n}(y_{i}|\mu _{j},\sigma _{j}^{2}),$ $%
j=0,\pm 1,\pm 2,\ldots $, each stage of the MCMC\ algorithm proceeds by
sampling from the following full conditional posterior densities:

\begin{enumerate}
\item $\pi (\mu _{j}|\cdots )=$ \textrm{n}$\left( \mu _{j}\left\vert \dfrac{%
\mu _{\mu j}\sigma _{j}^{2}+n_{j}\sigma _{\mu j}^{2}\overline{y}_{j}}{\sigma
_{j}^{2}+n_{j}\sigma _{\mu j}^{2}},\dfrac{\sigma _{j}^{2}\sigma _{\mu j}^{2}%
}{\sigma _{j}^{2}+n_{j}\sigma _{\mu j}^{2}}\right. \right) ,$ for $j=0,\pm
1,\ldots ,\pm N_{\max }$, with $n_{j}=\tsum\limits_{i:z_{i}=j}1$, $\overline{%
y}_{j}=\tfrac{1}{n_{j}}\tsum\limits_{i:z_{i}=j}y_{i},$ $N_{\max
}=\max_{i}[\max_{j}\mathbf{1}(u_{i}<\xi _{j})|j|]$, given $n$\ independent
uniform random draws $u_{i}\sim $ \textrm{u}$(0,\xi _{|d_{i}|})$, $%
i=1,\ldots ,n$;

\item $\pi (\sigma _{j}^{-2}|\cdots )=$ \textrm{ga}$\left( \sigma
_{j}^{-2}\left\vert \alpha _{\sigma j}+\tfrac{1}{2}n_{j},\beta _{\sigma j}+%
\tfrac{1}{2}\tsum\limits_{i:z_{i}=j}(y_{i}-\mu _{j})^{2}\right. \right) $,
for $j=0,\ldots ,\pm N_{\max }$;

\item $\Pr (d_{i}=j|\cdots )\propto \mathbf{1}(u_{i}<\xi _{j})\xi _{j}^{-1}%
\mathrm{n}(y_{i}|\mu _{j},\sigma _{j}^{2})P(z(\mathbf{x}_{i})\in A_{j}),$
for $j=0,\ldots ,\pm N_{\max }$\ and for $i=1,\ldots ,n$, where $P(z(\mathbf{%
x}_{i})\in A_{j})=\tint\nolimits_{j-1}^{j}\mathrm{n}\left( z(\mathbf{x}%
_{i})|\eta _{i}^{\ast },\psi _{ii}^{-1}\right) \mathrm{d}z,$ and $\eta
_{i}^{\ast }=(1,\mathbf{x}_{i}^{\intercal })\boldsymbol{\beta }%
+\tsum\limits_{l\neq i}(-\psi _{il}/\psi _{ii})(z(\mathbf{x}_{l})-(1,\mathbf{%
x}_{l}^{\intercal })\boldsymbol{\beta }),$ given the precision matrix, $\Psi
_{\boldsymbol{\phi }}^{(n)}=(\sigma _{\mathcal{C}}^{2}\mathcal{C}_{%
\boldsymbol{\phi }}\mathbf{(\mathbf{x}}_{i},\mathbf{\mathbf{x}}%
_{l}))_{n\times n}^{-1}=(\psi _{il})_{n\times n}$;

\item $\pi (z(\mathbf{x}_{i})|\cdots )\propto \mathbf{1}(z(\mathbf{x}%
_{i})\in A_{d_{i}}=(d_{i}-1,d_{i}])\mathrm{n}(z(\mathbf{x}_{i})|\eta
_{i}^{\ast },\psi _{ii}^{-1}),$ for $i=1,\ldots ,n$;

\item $\pi (\boldsymbol{\beta }|\cdots )=\mathrm{n}(\boldsymbol{\beta }|%
\mathbf{m}_{\boldsymbol{\beta }}^{\ast },\phi _{1}\mathbf{V}_{\boldsymbol{%
\beta }}^{\ast }),$ given $\mathbf{V}_{\boldsymbol{\beta }}^{\ast }=(\mathbf{%
V}_{\boldsymbol{\beta }}^{-1}+\mathbf{X^{\intercal }\Psi }_{\boldsymbol{\phi 
}}^{(n)}\mathbf{X)}^{-1}$ and $\mathbf{m}_{\boldsymbol{\beta }}^{\ast }=%
\mathbf{V}_{\boldsymbol{\beta }}^{\ast }(\mathbf{V}_{\boldsymbol{\beta }%
}^{-1}\mathbf{m}_{\boldsymbol{\beta }}+\mathbf{X}^{\intercal }\Psi _{%
\boldsymbol{\phi }}^{(n)}\mathbf{z}),$ where $\mathbf{z}_{n}=(z(\mathbf{x}%
_{1}),\ldots ,z(\mathbf{x}_{n}))^{\intercal }$;

\item $\pi (\sigma _{\mathcal{C}}^{2}|\cdots )=\mathrm{ga}(\sigma _{\mathcal{%
C}}^{-2}|a_{\mathcal{C}}+n/2,b_{\phi }+\{\mathbf{m}_{\boldsymbol{\beta }%
}^{\intercal }\mathbf{V}_{\boldsymbol{\beta }}^{-1}\mathbf{m}_{\boldsymbol{%
\beta }}-\mathbf{z}_{n}^{\mathbf{\intercal }}\Psi _{\boldsymbol{\phi }}^{(n)}%
\mathbf{z}_{n}\mathbf{-(m}_{\boldsymbol{\beta }}^{\ast })^{\intercal }(%
\mathbf{V}_{\boldsymbol{\beta }}^{\ast })^{-1}\mathbf{m}_{\boldsymbol{\beta }%
}^{\ast }\}/2)$;

\item $\pi (\boldsymbol{\phi }|\cdots )\propto \mathrm{n}(z(\mathbf{x}%
_{1}),\ldots ,z(\mathbf{x}_{n})|\mathbf{X}_{n}\boldsymbol{\beta },\sigma _{%
\mathcal{C}}^{2}(\mathcal{C}_{\boldsymbol{\phi }}(\mathbf{x}_{i},\mathbf{x}%
_{l}))_{n\times n})\pi (\boldsymbol{\phi })$;

\item $f(y^{\mathrm{pred}}|\mathbf{x},\cdots )\propto \mathrm{n}(y|\mu
_{j},\sigma _{j}^{2})\mathbf{1}(z(\mathbf{x})\in A_{j})\mathrm{n}(z(\mathbf{x%
})|\mu ^{\ast }(\mathbf{x}),\sigma _{\boldsymbol{\phi }}^{\ast }(\mathbf{x}%
)) $ for each covariate input $\mathbf{x}$ of interest, where $\mathbf{%
\sigma }_{\boldsymbol{\phi }}(\mathbf{x})=\sigma _{\mathcal{C}}^{2}(\mathcal{%
C}_{\boldsymbol{\phi }}(\mathbf{x},\mathbf{x}_{1}),\ldots ,\mathcal{C}_{%
\boldsymbol{\phi }}(\mathbf{x},\mathbf{x}_{n}))^{\intercal }$, $\mu ^{\ast }(%
\mathbf{x})=(1,\mathbf{x}^{\mathbf{\intercal }})\boldsymbol{\beta }+\mathbf{%
\sigma }_{\boldsymbol{\phi }}(\mathbf{x})^{\intercal }\Psi _{\boldsymbol{%
\phi }}^{(n)}(\mathbf{z}_{n}-\mathbf{X}_{1n}\boldsymbol{\beta }),$ and $%
\sigma _{\boldsymbol{\phi }}^{\ast }(\mathbf{x})=\sigma _{\mathcal{C}}^{2}%
\mathcal{C}_{\boldsymbol{\phi }}(\mathbf{x},\mathbf{x})-\mathbf{\sigma }_{%
\boldsymbol{\phi }}(\mathbf{x})^{\intercal }\Psi _{\boldsymbol{\phi }}^{(n)}%
\mathbf{\sigma }_{\boldsymbol{\phi }}(\mathbf{x}).$\noindent
\end{enumerate}

The full conditionals in Steps 1-6 and 8 can be sampled directly, using
\noindent standard theory for Bayesian linear models, GP\ models, and
standard methods for sampling truncated normal distributions (e.g., O'Hagan
\&\ Forster, 2004\nocite{OhaganForster04}; Damien \&\ Walker, 2001\nocite%
{DamienWalker01}). The full conditional in Step 7 can be sampled using a
Metropolis-Hastings or another rejection-sampling algorithm, if necessary.
Step 8 of the MCMC algorithm provides samples from the posterior predictive
density $f_{n}(y|\mathbf{x})$ of the regression model. The full 8-step
sampling algorithm is repeated a large number $S$ of times, to construct a
discrete-time Harris ergodic Markov chain $\{\boldsymbol{\zeta }^{(s)}=(%
\boldsymbol{\mu },\boldsymbol{\sigma }^{2},\boldsymbol{\beta },\sigma _{%
\mathcal{C}}^{2},\boldsymbol{\phi })^{(s)}\}_{s=1}^{S}$ having a posterior
distribution $\Pi (\boldsymbol{\zeta }|\mathcal{D}_{n})$ as its stationary
distribution, provided a proper prior $\Pi (\boldsymbol{\zeta })$. We have
written MATLAB\ (2012, The MathWorks, Natick, MA) code that implements the
MCMC sampling algorithm. Standard methods can be used to check whether the
MCMC\ algorithm has generated a sufficiently-large number of samples from
the model's posterior distribution. Specifically, given MCMC\ samples $\{%
\boldsymbol{\zeta }^{(s)}\}_{s=1}^{S}$ generated by the algorithm,
univariate trace plots of these samples can be used to evaluate the mixing
of the chain (i.e., the degree to which the chain explores the support of
the posterior distribution). Also, for a posterior (moment or quantile)
estimate of any chosen scalar functional $\varphi (\boldsymbol{\zeta })$, a
Monte Carlo Confidence Interval (MCCI) can be computed via an applications
of a batch means method (for posterior moment estimates) or a subsampling
method (for posterior quantile estimate) applied to the MCMC samples $%
\{\varphi (\boldsymbol{\zeta }^{(s)})\}_{s=1}^{S}$.

Simple modifications of the MCMC\ algorithm and/or our model (Section \ref%
{The Regression Model section}) can be used to address other important data
analysis tasks:

\begin{itemize}
\item \underline{Multiple imputation of a censored dependent response $y_{i}$%
}: At each iteration of the MCMC algorithm, a plausible value of a dependent
response $y_{i}$ that is censored and known only to fall within an interval $%
(a_{y_{i}},b_{y_{i}}],$ is sampled from the full conditional posterior
predictive density $\pi (y|\mathbf{x}_{i},\cdots )$ $\propto $ $\mathrm{n}%
(y|\mu _{d_{i}},\sigma _{d_{i}}^{2})\mathbf{1}(y\in (a_{y_{i}},b_{y_{i}}])$,
and then is imputed as the updated value of $y_{i}$.

\item \underline{Discrete-valued dependent variable}: Our regression model
can be extended to handle ordinal discrete-valued dependent variable
responses, $y_{i}\in \{0,1,\ldots ,C_{i}^{\max }\geq 1\}$ ($i=1,\ldots ,n$),
by using instead probit kernels of the form $f(c|\boldsymbol{\theta }_{j})=$ 
$\tint_{\mathcal{A}(c)}$\textrm{n}$(y^{\ast }|\mu _{j},\sigma _{j}^{2})%
\mathrm{d}y^{\ast }$ ($j=0,\pm 1,\pm 2,\ldots $), for disjoint sets $%
\mathcal{A}(c)$ such that $\cup _{c=0}^{C}\mathcal{A}(c)=%
%TCIMACRO{\U{211d} }%
%BeginExpansion
\mathbb{R}
%EndExpansion
.$ In this case, we would add a step to the existing MCMC\ algorithm, to
sample from the full conditional posterior density of the latent variables $%
\pi (y_{i}^{\ast }|\mathbf{x}_{i},\cdots )\varpropto $ $\mathrm{n}%
(y_{i}^{\ast }|\mu _{d_{i}},\sigma _{d_{i}}^{2})\mathbf{1}(y_{i}^{\ast }\in 
\mathcal{A}(y_{i}))$, $i=1,\ldots ,n$. Then all the other steps of the
original MCMC\ algorithm proceeds with the current state of the latent
variables $y_{i}^{\ast }$ ($i=1,\ldots ,n$) instead of the $y_{i}$ ($%
i=1,\ldots ,n$).

\item \underline{Spatio-temporal setting}: In such a setting, we may specify
the covariance function $\sigma _{\mathcal{C}}^{2}\mathcal{C}_{\boldsymbol{%
\phi }}(\mathbf{x},\mathbf{x}^{\prime })=\sigma _{\mathcal{C}}^{2}\mathcal{C}%
_{\boldsymbol{\phi }_{1}}(\underline{\mathbf{x}},\underline{\mathbf{x}}%
^{\prime })\mathcal{C}_{\boldsymbol{\phi }_{2}}(\mathbf{s},t;\mathbf{s}%
^{\prime },t^{\prime })$, given covariates $\underline{\mathbf{x}}\mathbf{,}$
spatial locations $\mathbf{s\in 
%TCIMACRO{\U{211d} }%
%BeginExpansion
\mathbb{R}
%EndExpansion
}^{K}$, and time $t\in \mathbf{%
%TCIMACRO{\U{211d} }%
%BeginExpansion
\mathbb{R}
%EndExpansion
}$, where $\mathcal{C}_{\boldsymbol{\phi }_{2}}(\cdot ,\cdot )$ denotes a
correlation function for non-separable space and time effects (Gneiting \&\
Guttorp, 2010\nocite{GneitingGuttorp10b}). For example, the covariance
function:%
\begin{eqnarray*}
\sigma _{\mathcal{C}}^{2}\mathcal{C}_{\boldsymbol{\phi }}(\mathbf{x},\mathbf{%
x}^{\prime }) &=&\sigma _{\mathcal{C}}^{2}\mathcal{C}_{\boldsymbol{\phi }%
_{1}}(\underline{\mathbf{x}},\underline{\mathbf{x}}^{\prime })\mathcal{C}_{%
\boldsymbol{\phi }_{2}}(\mathbf{s},t;\mathbf{s}^{\prime },t^{\prime }) \\
&=&\sigma _{\mathcal{C}}^{2}\exp (-.5||\underline{\mathbf{x}}-\underline{%
\mathbf{x}}^{\prime }||^{2})\exp \left( -.5\{||\mathbf{s}-\mathbf{s}^{\prime
}||^{2}/2+||t-t^{\prime }||^{2}/2\}\right) .
\end{eqnarray*}

\item \underline{Estimating $D_{\tau }(m)$}:\ For a given Bayesian model $m$%
, the estimate of the criterion $D(m)$ is obtained by $\widehat{D}(m)=\tfrac{%
1}{S}\tsum\nolimits_{i=1}^{n}\{y_{i}-y_{i}^{\mathrm{pred}(s)}\}^{2}$, given
posterior predictive samples $\{\{(y_{i}^{\mathrm{pred}(s)}|\mathbf{x}%
_{i},m)\}_{i=1}^{n}\}_{s=1}^{S}$.
\end{itemize}

\bigskip

\noindent \begingroup%
%TCIMACRO{\TeXButton{TeX field}{\bf\Large}}%
%BeginExpansion
\bf\Large%
%EndExpansion
\noindent {\Large \noindent }Acknowledgements\endgroup

This research is supported by National Science Foundation research grant
SES-1156372, from the program in Methodology, Measurement, and Statistics.
This paper was presented, in part, at the invited session on Bayesian
nonparametrics, of the ERCIM Working Group conference on Computing and
Statistics, Oviedo, Spain, December, 1-3, 2012

\bibliographystyle{rss}
\bibliography{Karabatsos}
\newpage

%TCIMACRO{\TeXButton{B}{\begin{table}[H] \centering}}%
%BeginExpansion
\begin{table}[H] \centering%
%EndExpansion
\begin{tabular}{|l|ccc|}
\hline
\textbf{Generating} & \multicolumn{3}{c|}{$D(m)$} \\ \cline{2-4}
\textbf{Model:} & \textbf{GP} & \textbf{Indep} & \textbf{BART} \\ \hline
2-dimensional ($n=225$) & $5.0$ $(\pm .1)$ & $75.\,\allowbreak 2$ $(\pm .5)$
& $48.7$ $(\pm .6)$ \\ 
10-dimensional ($n=100$) & $2.9$ $(\pm .3)$ & $577.8$ $(\pm 7.4)$ & $140.2$ $%
(\pm 3.1)$ \\ 
Multimodal ($n=100$) & $2.3$ $(\pm .1)$ & $18.0$ $(\pm .4)$ & $2691.0$ $(\pm
9.1)$ \\ 
Multimodal ($n=225$) & $27.6$ $(\pm 2.5)$ & $316.9$ $(\pm 7.4)$ & $6415.4$ $%
(\pm 11.4)$ \\ \hline
\end{tabular}%
\caption{Results of the Simulation Study. Predictive accuracy of the
GP-based regression model, versus the independence model. (Each number in parentheses gives the corresponding 95 percent MCCI.)}%
\label{Sim Table}%
%TCIMACRO{\TeXButton{E}{\end{table}}}%
%BeginExpansion
\end{table}%
%EndExpansion

\begin{center}
\bigskip 

\textbf{FIGURE CAPTIONS}
\end{center}

\underline{Figure 1}. The mixture weights $\omega _{j}(\mathbf{x})$ and
corresponding predictive density $f(y|\mathbf{x})$ of the model. The figure
assumes $\eta (x_{1})=-.30$, $\eta (x_{2})=.21$, $\eta (x_{3})=4.8$, and the
covariance function $\sigma (x,x^{\prime })=\sigma _{C}^{2}\exp
(-.5||x-x^{\prime }||^{2})$.

\underline{Figure 2}. For the GP\ model, the posterior predictive mean of
the change score (solid line) plus/minus 2 times the posterior predictive
variance (dashed lines).

\underline{Figure 3}. For the GP\ model, the posterior predictive density
estimates, given a range of pretest scores, and conditional on uic=1 and
lmt140=1.

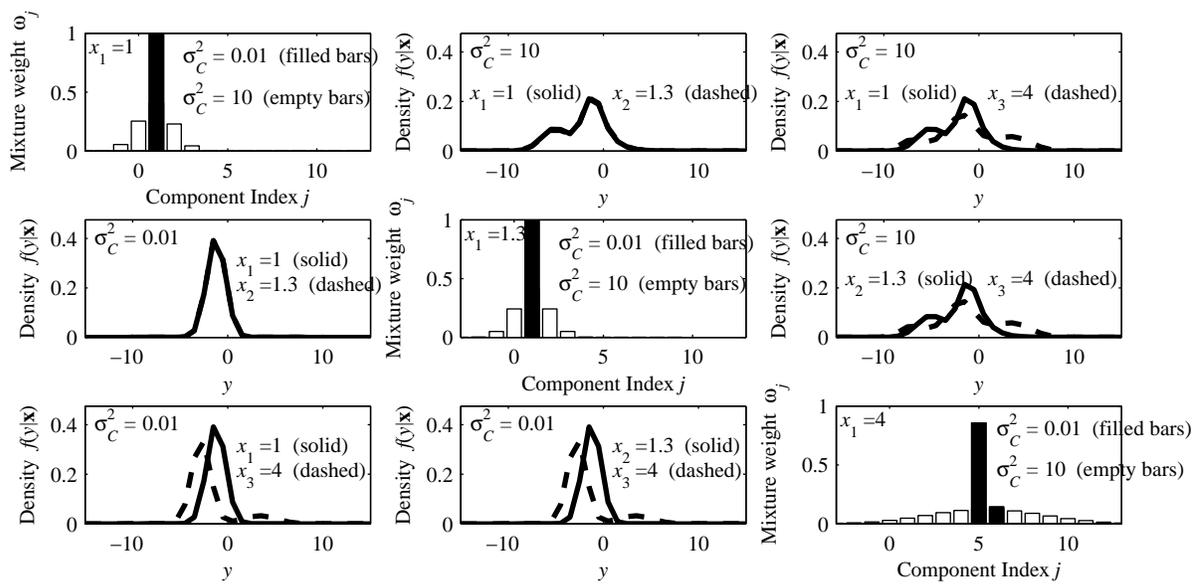
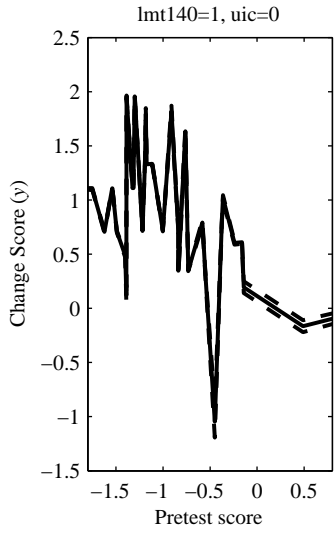
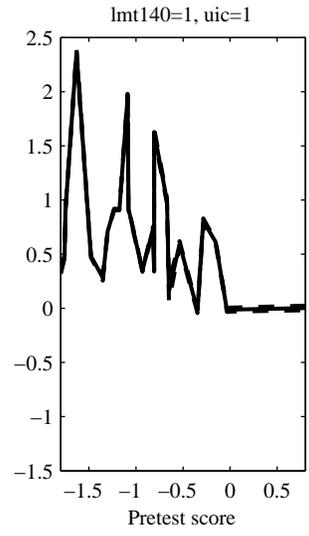
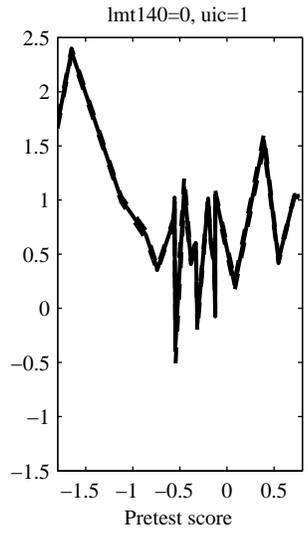
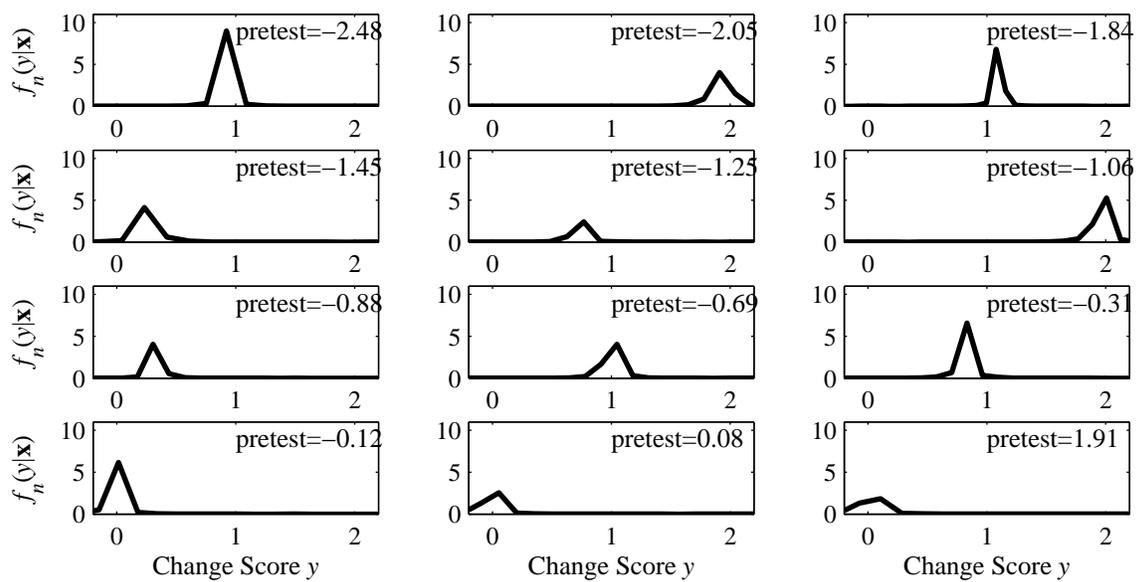
\end{document}